\documentclass[twocolumn,aps,prd,amsmath,amssymb,nofootinbib]{revtex4-1}

\usepackage{graphicx,color,bm,enumerate}

\newcommand{\beq}{\begin{equation}}
\newcommand{\eeq}{\end{equation}}
\newcommand{\bea}{\begin{eqnarray}}
\newcommand{\eea}{\end{eqnarray}}
\newcommand{\amp}{&}

\newcommand{\mpl}{M_{\rm Pl}}
\newcommand{\sca}{\chi}

\begin{document}

\title{Constraints on Gravitation from Causality and Quantum Consistency}

\author{Mark P.~Hertzberg}
\affiliation{Institute of Cosmology \& Department of Physics and Astronomy,\\
Tufts University, Medford, MA 02155, USA}

\date{\today}

\begin{abstract}
We examine the role of consistency with causality and quantum mechanics in determining the properties of gravitation. We begin by examining two different classes of interacting theories of massless spin 2 particles -- gravitons. One involves coupling the graviton with the lowest number of derivatives to matter, the other involves coupling the graviton with higher derivatives to matter, making use of the linearized Riemann tensor. The first class requires an infinite tower of terms for consistency, which is known to lead uniquely to general relativity. The second class only requires a finite number of terms for consistency, which appears as another class of theories of massless spin 2. We recap the causal consistency of general relativity and show how this fails in the second class for the special case of coupling to photons, exploiting related calculations in the literature. In a companion paper \cite{Sandora2016} this result is generalized to a much broader set of theories. Then, as a causal modification of general relativity, we add light scalar particles and recap the generic violation of universal free-fall they introduce and its quantum resolution. This leads to a discussion of a special type of scalar-tensor theory; the $F(\mathcal{R})$ models. We show that, unlike general relativity, these models do not possess the requisite counterterms to be consistent quantum effective field theories. Together this helps to remove some of the central assumptions made in deriving general relativity.
\end{abstract}

\maketitle

{\em Introduction}.---\hspace{-0.1cm}
General relativity is consistent with observations over a vast range of length scales. The $1/r^2$ force law has been tested down to fractions of a millimeter, while precisions tests of the relativistic theory have occurred on solar system scales, binary pulsars, and even the recent gravitational wave observations of merging binary black holes. On galactic and cosmological scales there is also agreement, though it does require the introduction of some as yet undiscovered form of dark matter and a small but nonzero amount of dark energy \cite{Riess:1998cb}.

The latter has provided some of the central motivations for considering alternatives to general relativity. It is difficult to understand why the vacuum energy is so small, despite there being known large contributions from massive particles running in loops, such as top quarks. Furthermore, the coincidence problem (why there is a comparable amount of matter and dark energy today), as well as the cosmological horizon, homogeneity, and flatness problems are also sometimes invoked as motivations. Also, there are a suite of difficulties in understanding general relativity as a quantum theory, including nonrenormalizability, trans-Planckian unitarity violation, black hole information paradox, and global issues associated with de Sitter space and eternal inflation. 

This range of primarily conceptual challenges, leads one to enquire just how inevitable general relativity is; whether theoretically consistent alternatives  exist. It is sometimes thought that indeed general relativity follows inevitably as the {\em unique} consistent theory of massless spin 2 particles at low energies. Following uniquely if one only assumes the Lorentz symmetry applied to the spin 2 degrees of freedom. That to deviate from general relativity requires either a violation of Lorentz symmetry, or the introduction of additional degrees of freedom. 

In this letter we clarify some aspects of this basic idea. Firstly, we point out that in fact classes of theories of massless spin 2 particles exist, which are Lorentz invariant and do not propagate new degrees of freedom. The most basic one involving the least number of derivatives, so called minimal coupling, and the others involving higher derivatives (the latter can be organized to not propagate any additional degrees of freedom). While the first leads to general relativity, the seconds appears as another class of theories of spin 2, and was earlier introduced in Ref.~\cite{Wald:1986bj}. We examine this second class in the special case of coupling to photons. We examine the propagation of photons in the second theory, showing there is superluminality (by exploiting related results in the literature) and forbidding a possible UV completion. In a companion paper \cite{Sandora2016} this idea is developed further and generalized to a much larger set of theories, including couplings to fermions and scalars, using a more systematic analysis. 

Secondly, we study conventional ways to modify general relativity by the addition of new light scalars. We emphasize that in the Standard Model of particle physics only the Lorentz symmetry is postulated and most couplings compatible with it are observed. Similarly, new scalars should generically come with many parameters leading to violation of the observed universality of free-fall. However, quantum effects typically remove the problem by making the scalars heavy. This leads to an examination of the so-called $F(\mathcal{R})$ models, which do have the property of immediately ensuring the universality of free-fall. We show that such theories, unlike general relativity, fail to have the appropriate set of counterterms in the domain of applications of these theories, and so they fail to be consistent quantum effective field theories.

{\em Massless Spin 2 Particles}.---\hspace{-0.1cm}
We are interested in constructing theories of massless spin 2 particles from the ground up. As is well known, the massless spin 2 unitary representation of the Lorentz group involves two helicities in 3+1 dimensions. It is useful to embed these two degrees of freedom into a symmetric tensor field in order to build a local theory. We shall denote this $h_{\mu\nu}$ and we will shortly discuss to what extent this may be interpreted as a metric field. Since $h_{\mu\nu}$ is a 10 component object, we need to remove 8 of the 10 degrees of freedom; 4 are removed by the introduction of constraints on $h_{\mu0}$, while another 4 can be removed by the introduction of an identification $h_{\mu\nu}\equiv h_{\mu\nu}+\partial_\mu\alpha_\nu+\partial_\nu\alpha_\mu$, where $\alpha_\mu$, the gauge function, parameterizes a family of physical equivalent representations of the same state $|\psi\rangle$. 

If one fixes the gauge, then under a Lorentz transformation $\Lambda^\rho_{\,\,\mu}$, the field $h_{\mu\nu}$ transforms as
\beq
h_{\mu\nu}\to \Lambda_{\,\,\mu}^\rho\,\Lambda^\sigma_{\,\,\nu}\,h_{\rho\sigma}+\partial_\mu\Omega_\nu+\partial_\nu\Omega_\mu
\eeq
where $\Omega_\mu$ depends on the gauge choice and $\Lambda^\rho_{\,\,\mu}$. Since $h_{\mu\nu}$ is evidently {\em not} a Lorentz tensor, it is generally very difficult to construct a Lorentz invariant interacting theory when we couple $h_{\mu\nu}$ to matter. There does exist, however, a manifestly gauge invariant and indeed Lorentz covariant 4-tensor we can construct out of derivatives of $h_{\mu\nu}$; the linearized Riemann tensor
\beq
R^{(L)}_{\mu\nu\rho\sigma}\equiv{1\over2}\left(\partial_\rho\partial_\nu h_{\mu\sigma}+\partial_\sigma\partial_\mu h_{\nu\rho}-\partial_\sigma\partial_\nu h_{\mu\rho}-\partial_\rho\partial_\mu h_{\nu\sigma} \right)
\label{RiemannLinear}\eeq
which we will explicitly make use of in the upcoming ``Type II" theories.

The free theory of these spin 2 particles is associated with terms of the form $\sim(\partial h)^2$. By demanding Lorentz/gauge invariance, only a unique set of terms is allowed, up to boundary terms,  which is
\beq
\mathcal{L}_{kin}^{h} = {1\over2}(\partial h)^2-{1\over2}\partial h_{\mu\nu}\partial h^{\mu\nu}+\partial_\mu h^{\mu\nu}\partial_\nu h-\partial_\mu h^{\rho\sigma}\partial_\rho h^\mu_\sigma
\eeq
where $h\equiv h^\mu_{\,\,\mu}$ and we have scaled the coefficient of the first term to $1/2$ without loss of generality.

{\em Type I: Lowest Number of Derivatives}.---\hspace{-0.1cm}
The interaction that involves the least number of derivatives, and hence would be most relevant at large distances, is to attempt to couple $h_{\mu\nu}$ directly to matter as follows
\beq
\mathcal{L}_{int}=h_{\mu\nu}\,  \tau^{\mu\nu}_M
\eeq
where $\tau_M^{\mu\nu}$ is some symmetric tensor built out of the matter fields, whose properties we shall shortly identify. Evidently this term is not gauge invariant for a generic $\tau_M^{\mu\nu}$, which means the theory is not unitary as it would propagate the wrong number of degrees of freedom. Alternatively, one could try to define this term in a particular gauge to avoid additional degrees of freedom, but then one would find the term is not Lorentz invariant as $h_{\mu\nu}$ is not a proper Lorentz tensor.

It is easy to see that under a gauge transformation and an integration by parts, the problem is fixed by taking $\tau_M^{\mu\nu}$ to be conserved $\partial_\mu\tau_M^{\mu\nu}=0$. So $\tau_M^{\mu\nu}$ should be proportional to the matter energy-momentum tensor $T_M^{\mu\nu}$ as follows $\tau_M^{\mu\nu}=-{\kappa\over2}T_M^{\mu\nu}$, where $\kappa$ is a coupling. We are assured that this is conserved due to translation invariance (at least to leading order; more on this shortly). This immediately implies that all matter particles $\phi_i$ must couple universally to $h_{\mu\nu}$, with strength $\kappa_1=\kappa_2=\ldots=\kappa$, as only the {\em total} energy-momentum tensor is conserved for interacting particles. This implies the (weak) equivalence principle.

This ensures the theory is gauge invariant on-shell. To also be gauge invariant off-shell one must endow the matter fields with a gauge transformation rule, such as $\phi_i\to\phi_i+\kappa\,\alpha_\mu\partial^\mu\phi_i$ for scalars \cite{Schwartz}. However, the gauge invariance is only ensured to order $\kappa$, since the presence of this interaction means that energy and momentum will in general be exchanged between matter and gravitons, so $T_M^{\mu\nu}$ is no longer exactly conserved. This requires several fixes: (i) one must include the coupling of $h_{\mu\nu}$ to itself as gravitons carry energy and momentum, (ii) at higher order in $\kappa$ one must modify the gauge transformation rule for $h_{\mu\nu}$ to involve higher order corrections, (iii) the gauge transformation rule for matter must also involve higher order corrections, such as $\phi_i\to\phi_i+\kappa\,\alpha_\mu\partial^\mu\phi_i+{1\over2}\kappa^2\alpha_\mu\alpha_\nu\partial^\mu\partial^\nu\phi_i+\ldots$, and (iv) an infinite tower of interaction terms, in powers of $\kappa$, must be included with the schematic $``\,\,"$ form
\bea
\mathcal{L}_{int}\amp=\amp\kappa\left(-h_{\mu\nu}T^{\mu\nu}_M/2+``h(\partial h)^2"\right)\nonumber\\
\amp+\amp\kappa^2\left(``h^2\,\hat{T}_M"+``h^2(\partial h)^2"\right)+\ldots
\eea
where every term is determined uniquely in terms of $\kappa$, up to boundary terms and field redefinitions. Amazingly, this infinite series can be re-summed for any matter Lagrangian $\mathcal{L}_M(\phi_i,\eta_{\mu\nu})$ \cite{Deser:1969wk}, giving the Einstein-Hilbert action
\beq
S_I=\int\!d^4x\sqrt{-g}\left[{\mathcal{R}\over 16\pi G}+\mathcal{L}_M(\phi_i,g_{\mu\nu})\right]
\label{EHAction}\eeq
where $G\equiv\kappa^2/(16\pi)$, $g_{\mu\nu}\equiv \eta_{\mu\nu}+\kappa\,h_{\mu\nu}$, and the matter Lagrangian involves the lift to ``minimal coupling" $\eta_{\mu\nu}\to g_{\mu\nu}$ and $\partial_\mu\to\nabla_\nu$. The gauge invariance is lifted to the full diffeomorphism invariance $\phi_i(x^\mu)\to\phi_i(x^\mu+\kappa\,\alpha^\mu)$ and $\mathcal{R}$ is the fully nonlinear Ricci scalar. Now $g_{\mu\nu}$ inevitably has a geometric interpretation.

Hence Type I coupling leads uniquely to general relativity and all of its successes. This theory can even be quantized, in the low energy regime, with concrete quantum gravity predictions such as \cite{Donoghue:1993eb}; we shall return to this issue later. Furthermore, by including some small, but nonzero, vacuum energy in the matter Lagrangian this can even account for cosmic acceleration. This theory does lead to conceptual puzzles, such as the cosmological constant problem, black hole information paradox, etc, as mentioned in the introduction, but is in great agreement with observations.

{\em Type II: Higher Number of Derivatives}.---\hspace{-0.1cm}
Here we would like to describe a much bigger class of theories of massless spin 2 particles; this was introduced earlier in Ref.~\cite{Wald:1986bj} and some other work includes Ref.~\cite{Bai:2016hui}. By exploiting the gauge invariant object $R^{(L)}_{\mu\nu\rho\sigma}$ defined in eq.~(\ref{RiemannLinear}) we can immediately write down a manifestly gauge/Lorentz invariant class of theories by coupling it into essentially any four index object $\tilde\tau_M^{\mu\nu\rho\sigma}$ with interaction $\mathcal{L}_{int}=R^{(L)}_{\mu\nu\rho\sigma}\,\tilde\tau_M^{\mu\nu\rho\sigma}$ giving
\beq
S_{II}=\int\!d^4x\left[\mathcal{L}_{kin}^h+\mathcal{L}_M(\phi_i,\eta_{\mu\nu})+R^{(L)}_{\mu\nu\rho\sigma}\,\tilde\tau_M^{\mu\nu\rho\sigma}\right]
\eeq
Note that this object involves only a {\em finite} number of terms in powers of $h_{\mu\nu}$, unlike the Type I theory that involves an infinite tower of terms. (Both types are of course nonrenormalizable in 3+1 dimensions so higher order terms will be generated by quantum mechanics, which we shall return to later). Also note that for any matter Lagrangian $\mathcal{L}_M$, we are essentially free to choose any new independent object $\tilde\tau_M^{\mu\nu\rho\sigma}$ which may be built out of many new parameters, unlike in Type I where the interaction term comes uniquely specified by the matter Lagrangian through the minimal coupling procedure. We further note that in Type II we can have many different flavors of massless spin 2 particles, without any contradiction, while in Type I there can only be a single flavor.

As a concrete example of the interaction, if the matter involves $N$ vector fields with field strengths $F_{\mu\nu,i}$ we could choose $\tilde\tau_M^{\mu\nu\rho\sigma}$ to be
\beq
\tilde\tau_M^{\mu\nu\rho\sigma}=\sum_{i=1}^N\left[a_i\,F_i^{\mu\nu}F_i^{\rho\sigma}+b_i\,\eta^{\mu\rho}F_i^{\nu\beta}F^\sigma_{i\,\beta}+c_i\,\eta^{\mu\rho}\eta^{\nu\sigma}F^2_i\right]
\label{TauPhoton}\eeq
where the $a_i,\,b_i,\,c_i$ are couplings. In general some constraints may be placed on the relative sizes of the $a_i,\,b_i,\,c_i$ to avoid higher time derivatives and ghosts, however there is no requirement for the couplings to be universal. So $a_1\neq a_2\neq a_3\neq\ldots$ is permitted by the Lorentz symmetry. Hence such a theory does not imply the (weak) equivalence principle.

{\em Principles in Physics}.---\hspace{-0.1cm}
If we considered the equivalence principle to be another fundamental postulate, then this would suffice to reject this entire Type II class in favor of the very special Type I.  However, in this work we only take the Lorentz symmetry as a fundamental postulate, and the equivalence principle is to be derived rather than assumed.

In fact it is useful to put this point of view in a broader perspective. It is often suggested that modern particle physics is built out of various additional postulates, such as the ``gauge principle" or ``principle of minimal coupling". However, if we examine the structure of the Standard Model, in particular its symmetries, a different picture emerges: (i) Exact symmetries: CPT derives from locality and unitarity, while $SU(3)\times SU(2)\times U(1)$ gauge is derived as an identification to remove the unphysical components of fields associated with twelve spin 1 particles. (ii) Approximate symmetries: $U(1)_B$, $U(1)_{B-L}$ are derived as accidental. (iii) Asymmetries: $C$, $P$, $T$, chiral, scale, etc, are not derivable from Lorentz symmetry and are not realized in nature. So in the Standard Model of particle physics, only that which follows from the Lorentz symmetry (applied to unitary representations) is realized, and no additional postulates appear to be required. 

Furthermore, the global $U(1)$, associated with electric charge, derives from considering the analogous Type I theory of massless photons coupled to charged matter with the lowest number of derivatives as $A_\mu J^\mu_M$, which requires coupling to a very special conserved $J^\mu_M$ (just as $\tau^{\mu\nu}_M$ above). On the other hand, we can also consider an analogous Type II theory of massless photons coupled to neutral matter with a higher number of derivatives as 
\beq
\mathcal{L}_{int}=F_{\mu\nu}\,\tilde{J}^{\mu\nu}_M
\eeq 
which does not require anything special about $\tilde{J}^{\mu\nu}_M$ (just as $\tilde\tau_M^{\mu\nu\rho\sigma}$ above). In fact this describes the low energy effective theory of photons coupled to neutrinos with $\tilde{J}^{\mu\nu}_M\propto\bar\psi\,\sigma^{\mu\nu}\psi$, etc. So at the level of the effective theory, both Type I and Type II theories are realized in nature for photons coupled to matter. This begs a question for gravitation: why has nature chosen Type I and not the much larger class Type II of gravitons coupled to matter?

There is reason to think that causality provides a possible answer to this question. We examine this in greater detail and for a much broader class of models in a companion paper \cite{Sandora2016}. For now we illustrate the idea in the special case of coupling to photons.

{\em Causality in Type I}.---\hspace{-0.1cm}
It is well known that in general relativity with standard matter sources there is no problem with causality \cite{Visser:1998ua}. This can be seen as follows: Consider photons minimally coupled to gravity in Type I. In the geometrics optics limit, photons obey the null geodesic equation $k_\mu k_\nu g^{\mu\nu}=0$, where $k_\mu$ is the photon's 4-momentum. The leading deflection from null propagation $k^{(0)}_\mu$ on the Minkowski cone comes from expanding in powers of the gravitational coupling as $k_\mu=k^{(0)}_\mu+\kappa\,k^{(1)}_\mu+\ldots$ and $g^{\mu\nu}=\eta^{\mu\nu}-\kappa\,h^{\mu\nu}+\ldots$, giving $k_\mu k_\mu \eta^{\mu\nu}\approx \kappa\,k_\mu^{(0)} k_\nu^{(0)}h^{\mu\nu}$. Then using the linearized Einstein equations in the Lorenz gauge $\Box \bar{h}^{\mu\nu}=-\kappa\,T_M^{\mu\nu}$ and solving for a particular solution gives \cite{Visser:1998ua}
\beq
k_\mu k_\nu \eta^{\mu\nu}\approx 4\,G\!\int\!d^3x'{k_\mu^{(0)}k_\nu^{(0)}T_M^{\mu\nu}\!({\bf x}',t_R)\over|{\bf x}-{\bf x}'|}
\eeq
which clearly satisfies $k_\mu k_\mu \eta^{\mu\nu}\geq0$ for any matter $T_M^{\mu\nu}$ that satisfies null-energy condition. Hence light stays inside the Minkowski cone and slows down in accord with the Shapiro time delay.

{\em Causality in Type II}.---\hspace{-0.1cm}
Let us focus on the case of the four index object $\tilde\tau_M^{\mu\nu\rho\sigma}$ given in eq.~(\ref{TauPhoton}) and focus on a single species; the photon. The modified Maxwell equations for a region of space-time that is Ricci flat is
\beq
\partial_\mu F^\mu_{\,\,\nu}-4\,a\,R^{(L)}_{\mu\nu\rho\sigma}\partial^\mu F^{\rho\sigma}=0
\eeq
Note that this equation is {\em exact}, the derivatives in Type II are just ordinary, not covariant, derivatives.

In the geometric optics limit, the leading deflection from null propagation on the Minkowski cone is \cite{Shore:2003zc}
\beq
k_\mu k_\nu \eta^{\mu\nu} \approx - 8\,a\,R_{\mu\nu\rho\sigma}^{(L)}k^\mu k^\rho\,\epsilon^\nu_p\epsilon^\sigma_p \Big{|}_{(0)}
\eeq
where all terms on the right hand side are evaluated for free null propagation. Here $\epsilon_p^\nu$ is the photon's polarization unit vector with $p=1,2$ for each of the two modes. For an appropriate choice of polarization and direction of propagation, relative to the gravitational field, one can arrange for $k_\mu k_\nu\eta^{\mu\nu}<0$ going outside the Minkowski cone. Since these theories are manifestly Lorentz invariant, this leads to problems with causality. This idea is greatly generalized in a companion paper \cite{Sandora2016}.

In a related context of QED, minimally coupled to gravity, it is known that one can integrate out the electron and generate terms of the form (\ref{TauPhoton}) (with the nonlinear Riemann tensor) with coefficients $a,\,b,\,c\sim \alpha\,\kappa/m_e^2$ \cite{Drummond:1979pp}. However this does not produce superluminality since the leading Shapiro time delay dominates over this effect in its domain of applicability. Related ideas appear in the context of string theory \cite{Camanho:2014apa}. However, in the context of this new class of spin 2 theories, where this is the leading interaction, superluminality appears unavoidable.

{\em Additional Fields}.---\hspace{-0.1cm}
There does exist a manifestly causal way to modify gravitation. This involves the introduction of additional degrees of freedom. Since fermions do not mediate long range forces, and vectors (with minimal coupling) have sources that tend to neutralize, we focus on the remaining case of adding light scalars. 

Usually in the literature a scalar $\sca$ is added that is taken to couple universally to the trace of the matter energy-momentum tensor at leading order as
\beq
S = S_I+\int d^4x\sqrt{-g}\left[{1\over2}(\partial\chi)^2+\gamma\,\sca\,g_{\mu\nu}T^{\mu\nu}_M+\ldots\right]
\label{PhiTraceT}\eeq
where $\gamma$ is a coupling. Here the universal coupling is inserted to be compatible with the (weak) equivalence principle. But as described earlier in this letter, the Standard Model does not give any reason to utilize any principle beyond that of just the Lorentz symmetry. Since $\sca$ is a gauge singlet scalar, we could take any gauge invariant term in the Standard Model $\mathcal{L}_{SM,j}$ and multiply it by $\sca$
\beq
S = S_I+\int d^4x\sqrt{-g}\left[{1\over2}(\partial\chi)^2+\sca\sum_j\gamma_j\,\mathcal{L}_{SM,j}+\ldots\right]
\label{generic}\eeq
where $\gamma_j$ are arbitrary couplings, and obtain a Lorentz invariant theory. 

So to assume the form (\ref{PhiTraceT}) is to tune the theory to be compatible with tests of the universality of free-fall, which have constrained $|\Delta a/a|<10^{-13}$ \cite{Schlamminger:2007ht}. One may appeal to technical naturalness to justify such universal couplings \cite{Hui:2010dn}, or to link $\sca$ to the dilaton of a spontaneously broken scale symmetry \cite{ArmendarizPicon:2011ys}, or to special fields associated with extra dimensions. However, generic scalars beyond the Standard Model do not have this feature of universal coupling. Instead Lorentz symmetry suggests the non-universal (\ref{generic}) is much more generic.

This poses a challenge to deriving the (weak) equivalence principle. However, if we take quantum effects into account, then a generic scalar $\sca$ will pick up a mass from Standard Model particles running in a loop of the form $\Delta m_\sca\sim\gamma\,\Lambda_{UV}^2/(4\pi)$ (or $\Delta m_\sca\sim\gamma\,m_{SM}\Lambda_{UV}/(4\pi)$), using a hard UV cutoff on the loop integral $\Lambda_{UV}$. For $\gamma\gtrsim\kappa$, and unless the cutoff is extremely low, the scalar $\sca$ will be typically heavy and unable to mediate long range forces, so general relativity is recovered at large distances.

{\em F($\mathcal{R}$) Gravity: Classical Treatment}.---\hspace{-0.1cm}
A popular framework that is both causal and enforces the universality of free-fall are the so-called $F(\mathcal{R})$ models. Here the Einstein-Hilbert action eq.~(\ref{EHAction}) is modified as $\mathcal{R}\to F(\mathcal{R})$. Note that inside any nonlinear function $F(\mathcal{R})$ are higher derivatives. This can be seen by expanding around a Minkowski background obtaining $\mathcal{R}=\kappa(\partial_\mu\partial_\nu h^{\mu\nu}-\Box h)+\mathcal{O}(\kappa^2 h(\partial h)^2)$. The $\sim\kappa\,\partial^2 h$ terms here only lead to a total derivative in the action if $F(\mathcal{R})$ is linear in $\mathcal{R}$, but for nonlinear $F(\mathcal{R})$ these higher derivatives have consequences. At the classical level, these consequences can be captured by the introduction of a scalar $\sca$ with action
\bea
S_\sca=\int\!d^4x\sqrt{-g}\Big{[}{\mathcal{R}\over 16\pi G}\amp+\amp\mathcal{L}_M(\phi_i,g_{\mu\nu}f(\sca))\nonumber\\
\amp+\amp{1\over2}(\partial\sca)^2-V(\sca)\Big{]}
\label{FRAction}\eea
where $f(\sca)$ and $V(\sca)$ are functions that depend on the choice of $F$. Note that $\sca$ couples to matter in a universal way through the single function $g_{\mu\nu}f(\sca)$, satisfying the (weak) equivalence principle.

A popular example is $F(\mathcal{R})=\mathcal{R}+\zeta\,\mathcal{R}^2/\mpl^2$ (with $\mpl\equiv1/\sqrt{16\pi G}$ and $\zeta\gg1$) which is a model of inflation \cite{Starobinsky:1980te}. 
Here the classically equivalent scalar $\chi$ plays the role of the inflaton. Its potential turns out to be
\beq
V(\sca)={\mpl^4\over4\,\zeta}\left(1-\exp\left(-{\sca\over\sqrt{3}\,\mpl}\right)\right)^2
\eeq
For large field values $\sca\gg\mpl$ the potential is exponentially flat and inflation takes place. One computes correlation functions of the scalar mode of the form $\langle \mbox{BD}|\sca(x)\sca(y)|\mbox{BD}\rangle$ (where $|\mbox{BD}\rangle$ is the Bunch-Davies vacuum) to obtain an approximately scale invariant spectrum of density perturbations with small red tilt $n_s-1\approx -2/N_e\sim-0.04$ and tensor-to-scalar ratio $r\approx 12/N_e^2\sim0.004$. These predictions are compatible with recent data \cite{Hinshaw:2012aka}. Similarly there exist many popular models of dark energy associated with various choices of $F(\mathcal{R})$ \cite{Sotiriou:2008rp}.

{\em F($\mathcal{R}$) Gravity: Quantum Treatment}.---\hspace{-0.1cm}
Here we would like to examine $F(\mathcal{R})$ gravity as an effective field theory. To begin, let's return to the Einstein-Hilbert action eq.~(\ref{EHAction}) and try to study it as a quantum theory. The quantum partition function is
\beq
Z_I=\int\prod_p\mathcal{D}h_p\!\prod_i\mathcal{D}\phi_i\,e^{iS_I[h_p,\,\phi_i]/\hbar}
\eeq
where the first measure of the path integral is over the {\em two} modes of the graviton labelled $p=1,2$ and the second measure is over the matter fields. In practice there are various complications associated with gauge fixing and constraints, but this is the formal structure. In principle this allows one to compute various correlation functions such as $\langle h_p(x)\,h_{p'}(y)\rangle$, $\langle\phi_i(x)\,\phi_j(y)\rangle$, etc. If we perform the path integral partially by integrating down to some scale $\Lambda$, we will generate a new Wilsonian effective action including corrections such as
\bea
S_{I,eff}=S_I(\Lambda)+\!\int\! d^4x\sqrt{-g}\Big{[}\amp\amp c_1(\Lambda)\,\mathcal{R}^2+c_2(\Lambda)\,\mathcal{R}_{\mu\nu}\mathcal{R}^{\mu\nu}\nonumber\\
+c_3(\amp\amp\Lambda)\,\mathcal{R}_{\mu\nu\rho\sigma}\mathcal{R}^{\mu\nu\rho\sigma}+\ldots\Big{]}\,\,
\label{EffAction}\eea
(plus some non-local terms, etc.) These additional terms are required as counterterms to cancel divergences associated with graviton loops. As long as we focus on only the physical degrees of freedom, such as the two modes of the graviton, we can use this effective theory to compute quantum effects. (In fact quantization of the leading Einstein-Hilbert term $S_I$ already gives rise to long range corrections to gravitation; see Refs.~\cite{Donoghue:1993eb}).

Note that the effective Lagrangian in eq.~(\ref{EffAction}) formally involves higher derivatives due to the presence of the terms $\mathcal{R}^2$, etc. Furthermore the presence of these types of terms might, at first sight, seem to justify the kind of $F(\mathcal{R})$ actions that we wrote above. However, it is essential to not use these higher derivative terms incorrectly; the original path integral is only defined with a measure for the two modes of the graviton and the matter fields. The measure does {\em not} include integration over additional degrees of freedom, such as a scalar $\sca$. Instead the path integral forces these to be spurious additional degrees of freedom; they can never be external and on-shell.

By contrast, the $F(\mathcal{R})$ models, as applied to inflation and dark energy etc, explicitly make use of the additional scalar $\sca$. This scalar is given its own dynamics, its own phase space, and its own independent set of fluctuations that are used as the source of density perturbations. This means the $F(\mathcal{R})$ models are disconnected from a rigorous quantum treatment of the Einstein-Hilbert action. 

One might attempt to quantize $F(\mathcal{R})$ using the path integral. However, the path integral requires one to integrate over all the physical fields in the theory. So we would need to explicitly use the scalar $\sca$ and form
\beq
Z_\sca=\int\prod_p\mathcal{D}h_p\,\mathcal{D}\sca\prod_i\mathcal{D}\phi_i\,e^{iS_\sca[h_p,\,\sca,\,\phi_i]/\hbar}
\eeq
Again by integrating over high energy modes, we form a new type of Wilsonian effective action
\bea
S_{\sca,eff}=S_\sca(\Lambda)+\!\int\! d^4x\sqrt{-g}\Big{[}d_1(\Lambda)(\partial\sca)^4+\ldots\Big{]}\,\,
\label{EffActionChi}\eea
(plus generating $\mathcal{R}^2$ terms, etc, again).
We emphasize the presence of new counterterms, such as $(\partial\sca)^4$. These additional counterterms are required to cancel new divergences that arise from the ability to put the $\sca$ external and on-shell. It is very important to note that such additional terms {\em cannot} be put in the $F(\mathcal{R})$ form, and more generally, cannot put in the form of some function only of the metric when generic matter is included. Hence the $F(\mathcal{R})$ models, which exploit the dynamics of the additional degree of freedom, contain cut-off dependence in the quantum theory without the required counterterms to be a consistent quantum effective theory.

We note that in other formulations of gravity, such as Palatini, similar conclusions hold. Namely, in constructions in which there are no new degrees of freedom, then the theory can always be recast into the general relativity form with a collections of appropriate counter-terms. While, in constructions in which there is a new degree of freedom, it can only be self-consistently quantized by re-organization into the scalar-tensor form.

{\em Outlook}.---\hspace{-0.1cm}
The above arguments help toward deriving general relativity as the only consistent theory involving massless spin 2 at low energies: (i) Type II theories that utilize higher derivative couplings (but can avoid extra degrees of freedom) can lead to problems with causality; see Ref.~\cite{Sandora2016} for an extended analysis. (ii) Additional scalars, which would generically lead to non-universal free-fall, are typically expected to be heavy due to quantum effects. (iii) $F(\mathcal{R})$ models are not consistent quantum effective field theories. 

However, important puzzles remain, including understanding dark energy. The smallness of the vacuum energy within the framework of general relativity does have a candidate explanation by introducing many (heavy) scalars, leading to a potential with an exponentially large number of vacua. Though it is unclear how to formulate probabilities in this context. While the behavior of gravitation at the Planck scale requires further new physics.

\vspace{0.1cm}

{\em Acknowledgments}.---\hspace{-0.1cm}
I would like to thank Raphael Flauger, Jaume Garriga, Alan Guth, David Kaiser, Juan Maldacena, McCullen Sandora, and Mark Trodden for helpful conversations. I would like to thank the Tufts Institute of Cosmology for support. An earlier version of this manuscript was presented at the 13th International Symposium on Cosmology and Particle Astrophysics (CosPA 2016)

{\em Email Correspondence.--- mark.hertzberg@tufts.edu}


\begin{thebibliography}{50}

% Upcoming paper
%\bibitem{Hertzberg:2017abn} 
\bibitem{Sandora2016}
  M.~P.~Hertzberg and M.~Sandora,
  %``General Relativity from Causality,''
  JHEP {\bf 1709}, 119 (2017)
%  doi:10.1007/JHEP09(2017)119
  [arXiv:1702.07720 [hep-th]].
  %%CITATION = doi:10.1007/JHEP09(2017)119;%%

% Data   Accelerating Universe   Riess:1998cb
\bibitem{Riess:1998cb} 
  A.~G.~Riess {\it et al.}  [Supernova Search Team Collaboration],
%  ``Observational evidence from supernovae for an accelerating universe and a cosmological constant,''
  Astron.\ J.\  {\bf 116}, 1009 (1998)
  [astro-ph/9805201].

%\cite{Wald:1986bj}
\bibitem{Wald:1986bj} 
  R.~M.~Wald,
  %``Spin-2 Fields and General Covariance,''
  Phys.\ Rev.\ D {\bf 33}, 3613 (1986).
 % doi:10.1103/PhysRevD.33.3613
  %%CITATION = doi:10.1103/PhysRevD.33.3613;%%

% GR derivation      Schwartz,Deser:1969wk
\bibitem{Schwartz}
M.~D.~Schwartz,
``Quantum Field Theory and the Standard Model,"
Cambridge University Press (2014).  
\bibitem{Deser:1969wk} 
  S.~Deser,
  %``Selfinteraction and gauge invariance,''
  Gen.\ Rel.\ Grav.\  {\bf 1}, 9 (1970)
%  doi:10.1007/BF00759198
  [gr-qc/0411023].
  %%CITATION = doi:10.1007/BF00759198;%%

% Quantum gravity predictions   Donoghue:1993eb,Ford:2015wls
\bibitem{Donoghue:1993eb} 
  J.~F.~Donoghue,
%  ``Leading quantum correction to the Newtonian potential,''
  Phys.\ Rev.\ Lett.\  {\bf 72}, 2996 (1994)
%  doi:10.1103/PhysRevLett.72.2996
  [gr-qc/9310024];
  %%CITATION = doi:10.1103/PhysRevLett.72.2996;%%
%\bibitem{Ford:2015wls} 
  L.~H.~Ford, M.~P.~Hertzberg and J.~Karouby,
%  ``Quantum Gravitational Force Between Polarizable Objects,''
  Phys.\ Rev.\ Lett.\  {\bf 116}, no. 15, 151301 (2016)
 % doi:10.1103/PhysRevLett.116.151301
  [arXiv:1512.07632 [hep-th]].
  %%CITATION = doi:10.1103/PhysRevLett.116.151301;%%

%\cite{Bai:2016hui}
\bibitem{Bai:2016hui} 
  D.~Bai and Y.~H.~Xing,
  %``Higher Derivative Theories for Interacting Massless Gravitons in Minkowski Spacetime,''
  Nucl.\ Phys.\ B {\bf 932}, 15 (2018)
%  doi:10.1016/j.nuclphysb.2018.05.009
  [arXiv:1610.00241 [hep-th]].
  %%CITATION = doi:10.1016/j.nuclphysb.2018.05.009;%%

% Causality in GR   Visser:1998ua,Adams:2006sv
\bibitem{Visser:1998ua} 
  M.~Visser, B.~Bassett and S.~Liberati,
%  ``Superluminal censorship,''
  Nucl.\ Phys.\ Proc.\ Suppl.\  {\bf 88}, 267 (2000)
%  doi:10.1016/S0920-5632(00)00782-9
  [gr-qc/9810026];
  %%CITATION = doi:10.1016/S0920-5632(00)00782-9;%%
%\bibitem{Adams:2006sv} 
  A.~Adams, N.~Arkani-Hamed, S.~Dubovsky, A.~Nicolis and R.~Rattazzi,
%  ``Causality, analyticity and an IR obstruction to UV completion,''
  JHEP {\bf 0610}, 014 (2006)
%  doi:10.1088/1126-6708/2006/10/014
  [hep-th/0602178].
  %%CITATION = doi:10.1088/1126-6708/2006/10/014;%%

% Causality in QED    Shore:2003zc
\bibitem{Shore:2003zc} 
  G.~M.~Shore,
%  ``Quantum gravitational optics,''
  Contemp.\ Phys.\  {\bf 44}, 503 (2003)
 % doi:10.1080/00107510310001617106
  [gr-qc/0304059].
  %%CITATION = doi:10.1080/00107510310001617106;%%
\bibitem{Drummond:1979pp} 
  I.~T.~Drummond and S.~J.~Hathrell,
%  ``QED Vacuum Polarization in a Background Gravitational Field and Its Effect on the Velocity of Photons,''
  Phys.\ Rev.\ D {\bf 22}, 343 (1980).
 % doi:10.1103/PhysRevD.22.343
  %%CITATION = doi:10.1103/PhysRevD.22.343;%%

% Causality in string theory    Camanho:2014apa
\bibitem{Camanho:2014apa} 
  X.~O.~Camanho, J.~D.~Edelstein, J.~Maldacena and A.~Zhiboedov,
%  ``Causality Constraints on Corrections to the Graviton Three-Point Coupling,''
  JHEP {\bf 1602}, 020 (2016)
%  doi:10.1007/JHEP02(2016)020
  [arXiv:1407.5597 [hep-th]].
  %%CITATION = doi:10.1007/JHEP02(2016)020;%%  

% Equivalence principle test    Schlamminger:2007ht
\bibitem{Schlamminger:2007ht} 
  S.~Schlamminger, K.-Y.~Choi, T.~A.~Wagner, J.~H.~Gundlach and E.~G.~Adelberger,
%  ``Test of the equivalence principle using a rotating torsion balance,''
  Phys.\ Rev.\ Lett.\  {\bf 100}, 041101 (2008)
 % doi:10.1103/PhysRevLett.100.041101
  [arXiv:0712.0607 [gr-qc]].
  %%CITATION = doi:10.1103/PhysRevLett.100.041101;%%

% Scalar tensor    Hui:2010dn, ArmendarizPicon:2011ys
\bibitem{Hui:2010dn} 
  L.~Hui and A.~Nicolis,
%  ``An Equivalence principle for scalar forces,''
  Phys.\ Rev.\ Lett.\  {\bf 105}, 231101 (2010)
%  doi:10.1103/PhysRevLett.105.231101
  [arXiv:1009.2520 [hep-th]].
  %%CITATION = doi:10.1103/PhysRevLett.105.231101;%%
\bibitem{ArmendarizPicon:2011ys} 
  C.~Armendariz-Picon and R.~Penco,
%  ``Quantum Equivalence Principle Violations in Scalar-Tensor Theories,''
  Phys.\ Rev.\ D {\bf 85}, 044052 (2012)
 % doi:10.1103/PhysRevD.85.044052
  [arXiv:1108.6028 [hep-th]].
  %%CITATION = doi:10.1103/PhysRevD.85.044052;%%

% F(R) inflation   Starobinsky:1980te
\bibitem{Starobinsky:1980te} 
  A.~A.~Starobinsky,
%  ``A New Type of Isotropic Cosmological Models Without Singularity,''
  Phys.\ Lett.\ B {\bf 91}, 99 (1980).
 % doi:10.1016/0370-2693(80)90670-X
  %%CITATION = doi:10.1016/0370-2693(80)90670-X;%%

% Data CMB    Hinshaw:2012aka, Ade:2013uln
\bibitem{Hinshaw:2012aka} 
  G.~Hinshaw {\it et al.}  [WMAP Collaboration],
%  ``Nine-Year Wilkinson Microwave Anisotropy Probe (WMAP) Observations: Cosmological Parameter Results,''
  arXiv:1212.5226 [astro-ph.CO];
% \bibitem{Ade:2013uln} 
  P.~A.~R.~Ade {\it et al.}  [Planck Collaboration],
%  ``Planck 2013 results. XXII. Constraints on inflation,''
  arXiv:1303.5082 [astro-ph.CO].
  
% F(R) gravity   Sotiriou:2008rp
\bibitem{Sotiriou:2008rp} 
  T.~P.~Sotiriou and V.~Faraoni,
 % ``f(R) Theories Of Gravity,''
  Rev.\ Mod.\ Phys.\  {\bf 82}, 451 (2010)
%  doi:10.1103/RevModPhys.82.451
  [arXiv:0805.1726 [gr-qc]].
  %%CITATION = doi:10.1103/RevModPhys.82.451;%%
  
\end{thebibliography}
\end{document}